\magnification=\magstep1
\hsize=15.5 truecm
\vsize=21.0 truecm
\leftskip=0.5 truecm                             
\topskip=3 truecm
\splittopskip=0 truecm
\parskip=0 pt plus 1 pt
\baselineskip=16.5 pt

\bigskip
\bigskip
\bigskip
\bigskip
\bigskip
\bigskip
\centerline{\bf Infinitesimally weak coupling, infinitely strong singularity} 
\centerline {\bf of the scattering potential}                            
\bigskip \bigskip
\centerline{ T. Dolinszky }
\bigskip
\centerline{  KFKI-RMKI,  H-1525 Budapest 114, POB 49, Hungary }
\centerline{  Email address: $<$Dolinszky@sgiserv.rmki.kfki.hu$>$}
\bigskip
\bigskip
\bigskip
\bigskip

\parindent 0pt
{\bf Abstract}

In scattering by singular potentials $g^2U(s;r)$, the coupling constant $g^2$
is  continuously decreased to zero while the stage $s$  of singularity
raised simultaneously  beyond all limits by some functional relation  
$F(g^2;s)=0$. In the extreme situation of this double limit, even the mere
existence of a nontrivial  physical scattering problem is questionable. By
iterating a pair of integral equations, the relevant   solution is developed
here in terms of wave functions   into a pair of convergent series, each of
which reduces  in the double limit $\{g^2\to 0;s\to\infty\}$   to a single
term calculable  by quadrature.    

\vfill
\eject 
\topskip=0 truecm

\ \ \ \ \  Just as in the regular case, also in the cases of repulsive
singular  potentials   $g^2U(s;r)$ the problem of scattering at extreme values
of the parameters  involved  is expected  to become solvable by some simple,
asymptotically exact  expressions. In this paper a review will be given  over
the main cases of varying the potential parameters contained either in
linear ($g^2$) or nonlinear (s) positions. The simplest  example is the 
increase, at invariable form factor, of the linear parameter beyond all
limits. This is the strong coupling limit, i.e.  $\{g^2\to\infty;s={\rm
fixed}\}$. For  singular potentials   this problem has already been solved by
a smooth version of the semiclassical approach [1,2]. The complementary
case is manifested by the variation of the nonlinear parameter $s$. The
notation to be applied  will ensure  that the stage
of singularity   should increase,  at fixed value of the coordinate $r$,  with
increasing values of $s$   to infinity.   The  asymptotical situation        
$\{g^2={\rm fixed};s\to\infty\}$ will be referred to here as the 
'supersingularity' limit. This  extreme  scattering problem was also attacked
by a  semiclassical procedure, which furnished   correct and simple results in
the  limit mentioned for a variety   of singular potentials [3].   
\smallskip    Nevertheless, within  our scheme of extreme scattering problems,
the most interesting point seems to be the $simultaneous$ variation of linear
and  nonlinear  potential  parameters. A particular example is  the double
limit $\{g^2\to 0;s\to\infty\}$. Such a problem is, of course, uniquely
specified only by  adding to the expression   $g^2U(s;r)$  as input
information a functional of the form  $F(g^2;s)=0$, which  governs the
interdependence of the parameters themselves.                         
\medskip                                               We are going to
scrutinize cases of such double limits via  combining  various classes of
potentials with different types of interdependence  between linear and
nonlinear pararameters. The underlying formalism is supplied by the approach
developed for solving the supersingularity problem in [3]. That argument will
be  briefly outlined  below.  A spinless particle is scattered by a central
singular potential at the energy $k^2$ in the  channel of index  $l$. To start
the discussion with, we   introduce a triad of auxiliary orbital angular
momenta  such as                        $$ \lambda_\epsilon^2(l)=(l+ {1\over
2})^2; \ \  \lambda_\tau^2(l)=l(l+1); \ \  \lambda^2(l)={1\over 2}[\lambda
_\epsilon^2(l)+\lambda_\tau^2(l)].\eqno(1) $$                       The
concept of the 'matching distance' $R$, into which the variable parameters
$g^2$ and $s$ will be lumped together, is introduced by the 'Master equation'
as                                $$ k^2R^2- g^2R^2U(s;R) -\lambda ^2=0, \ \ 
{\rm whence} \  \  R=R(g^2,s).\eqno(2) $$                          The
dimensionless radial coordinate  $t$, understood as                           
  $$ t={r\over R}; \ \ \ \  t^<_>1 \ \  {\rm if} \ \ r^<_>R,\eqno(3)  $$ 
works as an essential technical means. The exact radial Schroedinger equation
is recast in the new variable as                                            $$
 \big\{ { {\rm d^2} \over {\rm d} t^2}+ k^2R^2-g^2R^2U(s;Rt)- {l(l+1)\over
t^2} \big\} u^\pm(t)=0, \ \ [0\leq t].\eqno(4) $$              The regular
solution  $u^+(t)$ will  be represented by a pair of series, one for the
'exponential' (or $\epsilon$)  and another one for the 'trigonometric'  (or
$\tau$) region as follows:                                                    
  $$ u^+(t)= u^+_\epsilon(t)=\sum_{n=0}^{\infty}w_{\epsilon n}^+(t),\ \
[t<1];\eqno(5) $$    and                                                      
      $$ u^+(t)=u^+_\tau(t)=\sum _{m=0}^{\infty}w_{\tau m}^+(t), \ \
[t>1].\eqno(6) $$                                                           
The notation used here is resolved by the following set of identities:        
     $$ w_{\epsilon 0}^\pm (t)= \big( {k^2 \over K_\epsilon^2(t)}\big)^{1 \over
4} {\rm e}^{\pm R\int _1^t{\rm d}t'|K_\epsilon (t')|}; \eqno (7) $$           
   $$ w_{\tau 0}^\pm (t)= \big( {k^2\over K_\tau^2(t)}\big)^{1\over 4}\{C^\pm
\cos [R\int_1^t{\rm d}t'|K_\tau(t')|]+ S^\pm \sin [R\int_1^t {\rm
d}t'|K_\tau(t')|]\},\eqno(8) $$                                              
and the iteration scheme                                                     
      $$ w_{\epsilon n}^+(t) =\int_0^t{\rm d}t'G_\epsilon^+(t,t') 
\Delta_\epsilon(t') w_{\epsilon n-1}^+(t'), \ \ \ [t<1], \eqno(9) $$          
      $$ w_{\tau m}^+(t)= \int_1^t {\rm d}t'G_{\tau}^+(t,t')
\Delta_\tau (t')w_{\tau m-1}^+(t'), \ \ [t>1],\eqno(10) $$                   
where $n,m=1,2,3...$. The constants $C^+$  and  $S^+$ are specified by
requiring smooth matching at $t=1$, in principle  of  the infinite series (5)
and (6) themselves, in practice, however, of their higher order,  [N,M], 
cut-off approximations [3].  As regards $C^-$  and  $S^-$, they can be freely
chosen but for the condition  $C^+S^-\not= C^-S^+$. Furthermore, the local wave
number squares have been understood  as                                      
       $$ K_\gamma ^2(t)=\mp \{k^2-g^2U(s;Rt)-{\lambda_\gamma^2\over R^2t^2}\},
\ \ [\gamma= \big(^\epsilon _ \tau \big ),\ \  t^<_>1],\eqno(11) $$           
         while the resolvents are given by the definitions                    
       $$ G_\gamma^+(t,t')={1\over
d_\gamma^+}[w_\gamma^+(t)w_\gamma^-(t')- w_\gamma^-(t)w_\gamma ^+(t')],\ \
[\gamma=\epsilon,\tau], \eqno(12) $$ with  the Wronskians                    
       $$ d_\epsilon^+ = -2kR, \ \ \ \  d_\tau ^+= C^+S^--C^-S^+. \eqno(13 $$ 
Finally, the residual potentials are given in the respective regions  by      
       $$ \Delta _\gamma (t)= -{5\over 16}\big( {1\over K_\gamma^2(t)} {{\rm
d}K^2_\gamma(t)\over{\rm d}t}\big )^2+{1\over 4}{1\over K_\gamma ^2(t)}{{\rm
d}^2 K_\gamma ^2(t)\over {\rm d}t^2} -{\lambda _\gamma^2- l(l+1) \over 
t^2}.\eqno(14) $$              The series expansions (5)-(6) had been found
[1] absolutely convergent  whenever two integrals, $P_\gamma(t),\ \
[\gamma=\epsilon,\tau]$, are  bounded. That is to say, the convergence
criteria read                                $$ P_\gamma(t)\equiv 
R\int_{t_\gamma}^t{\rm{d}}t'|p_\gamma(t')|<c_\gamma<\infty,\ \ \ [\gamma=
\big(^\epsilon_\tau\big),\ \ t_\gamma =\big(^0_1\big),\ \ t^<_>1],\eqno(15) $$
where the concepts of the 'discriminants', namely                          
                                      $$ p_\gamma(t)\equiv
{\Delta_\gamma(t)\over RK_\gamma(t)}, \ \ \   [\gamma=\epsilon,\tau]
\eqno(16) $$       were introduced. 
  \medskip          The
potentials to be included in the discussion will occur as products  of 3
factors, such as a coupling constant $g^2$, a core factor $V_\epsilon(s;r)$
and a tail factor $V_\tau(r)$. The function $V_\epsilon(s;r)$ will be singular
at $r=0$ either exponentially or powerlaw in $r$, while $V_\tau(r)$ should 
decrease for $r\to \infty$ exponentially or powerlike. Owing to the
simultaneous variation of the parameters $g^2$ and $s$,  an everywhere extreme
and a locally extreme effect will face each other. Increasing values of $s$ 
should  correspond, by definition,  to raising stages of the singularity.  The
interparameter  relationship  $F(g^2;s)=0$  is to  appear  decomposed into a
pair of $R$-functions.  Out of them, the function $g^2(R)$ will be supplied as
input information while $s(R)$  introduced via  the Master equation (2). In
this way, the explicit presence of  both the linear and the nonlinear 
parameters can  be  eliminated from the  asymptotical formulas. The limit
$R\to \infty$ will be checked for  each potential class separately to recover
the double limit $g^2(R)\to 0; s(R)\to \infty$.    \bigskip Our choice of both
the  $R$-function $g^2(R)$ and  of the $r$-functions  $V_\epsilon(s;r)$ and
$V_\tau(r)$ is  either an  exponential (E) or a  powerlaw (P) dependence.  It
is therefore convenient  to refer to each of our  potential classes  by a
triad of the  capitals  $E$ or $P$, e.g.  EEE, EPP, PEE etc., in the order of 
  $g^2(R),V_\epsilon(r),V_\tau(r)$.                            
\medskip
\ \ \ \ \  {\bf Case EEE} 
\smallskip                         
This is, perhaps, the most interesting potential class in our
discussions. A rapidly decreasing coupling constant will  compete with a
rapidly raising  stage of the $r=0$  point singularity of the interaction.   
The set of formulae below leads from the  definition of the fixed parameter
form of the  potential up to proving fulfilment of the criteria for
convergence of the series (5)-(6) in the double limit considered. Accordingly,
we write                                              $$ g^2U(s;r)= {1\over
r_0^2}{\rm e}^{-{R\over r_0}}{\rm e}^{r_1s\over r}{\rm e}^{-{r\over r_2}},
\eqno(17) $$      with the  fixed positive constants $r_0, r_1, r_2$ and the 
variable  $s$. The  variation of the singularity parameter $s(R)$ is
governed by the Master equation (2), the exact form of which reads in this
potential class                                
         $$ {\rm{e}}^ {r_1s(R)\over R}=
(k^2-{\lambda^2\over R^2}) r^2_0 {\rm e}^{R({1\over r_0}+{1\over
r_2})}.\eqno(18) $$            
Considered as the definition of the  function
$s(R)$, this equation implies for $R\to\infty$ the order relationship 
O$\{s(R)\}$=O$\{{R^2\over r_0r_2}\}$. This verifies our expectation is
equivalent to the double limit   $\{g^2\to 0;s\to\infty\}$. Upon 
incorporating  Eq. (18) into the definition (11) one obtains                                                                                              
         $$ K_\gamma^2(t)\to \mp k^2 \{1-(k^2r_0^2)^{{1\over t}-1} {\rm
e}^{R[({1\over  r_0}+{1\over r_2}){1\over t}-({1\over r_0}+{1\over r_2}t)]
}-{\lambda_\gamma^2 \over k^2R^2t^2} \} \eqno(19) $$                          
for $R\to\infty$.  The notation is the same as the one used in Eq.(11). Hence
the discriminants  $p_\gamma(t)$ of the definition  (16) are  extracted, first
for the  region  $\epsilon$, as        $$  p_\epsilon(t) \to -{R\over
16k}\big[ \big({1\over r_0}+ {1\over r_2}\big) {1\over t^2} + {1\over
r_2}\big]^2 {\rm e}^{-{R\over 2}[ ({1\over r_0}({1\over t}-1) +{1\over
r_2}({1\over t}-t)]},  \ \ [R\to\infty] .\eqno(20) $$                         
  As to the region  $\tau$, the local wave number square (19) contains an
exponentially vanishing term, which greatly simplifies the formalism. Indeed,
one simply gets  in the long run                          $$ p_\tau
(t)\to-{3\lambda_\tau^2\over 2k^2R^2t^4},\ \  [R\to\infty].\eqno(21)$$      
Returning to the definitions (15)-(16), one concludes from  the relationship
(20) that the function  $P_\epsilon(t)$  is majorized in $t=(0,1)$ for the
case EEE  by $\Gamma(2)=1$.  As to the integral $P_\tau (t)$, it does not
involve in $t=(1,\infty)$ any singularity and so it also remains finite. 
\smallskip     The local wave number square (11)  contains  in the region
$\tau$ for       $R\to\infty$  in  each of our potential classes    
exponentially  vanishing interaction contributions only. These  become
asymptotically negligible in comparison to the energy and the centrifugal
term. As a consequence, expression (11) reduces  for $\gamma=\tau$ to the very
same formula (21), independently of the actual potential.  Thereby, the
convergence of the series (6) is guaranteed  in every case. The further
discussions  can be therefore restricted to the respective $\epsilon$ regions.       \medskip {
  
\medskip
\ \ \ \ \    {\bf Case EEP}
\smallskip                             
Within the potential classes to be
 included in the present discussions, the potential tail exerts
virtually no influence on the conflict between the vanishing  coupling
constant and the  increasing singularity of the core factor. Therefore, no
essential difference is expected between the cases EEE and EEP. The
present class of potentials is introduced as                                  
                                     $$ g^2U(s;r)={1\over r_0^2} {\rm
e}^{-{R\over r_0}}{\rm e}^{r_1s\over r} \big({r_2\over r_2+r}\big)^\sigma, \ \
[\sigma>8], \eqno (22) $$           where the quantities $r_0,r_1,r_2$ all are
positive and fixed against variation. The asymptotical  Master equation (2),
which specifies the function $s(R)$ for large values of $R$,  becomes in our
double limit                $$  {\rm e}^{r_1s(R)\over R}\to
\big(k^2-{\lambda^2\over R^2}\big)r_0^2{\rm e}^{R\over r_0}\big({R\over
r_2}\big)^\sigma,\ \ [R\to\infty], \eqno(23) $$    

whence one  extracts O$\{s(R)\}>$O$\{{R^2\over r_0r_1}\}$  for
$R\to\infty$.  The formulae (11) and  (23) combine then into  
           $$ K^2_\gamma(t)\to  \mp k^2 \big\{1 - [k^2r_0^2{\rm e}^
{R\over r_0}({R\over r_2})^\sigma]^{{1\over t}-1}-
{{\lambda_\gamma}^2\over k^2R^2t^2} \big\}. \eqno(24) $$                   
Hence one concludes for the exponential region  that by the identity  (16)     
                    $$ p_\epsilon(t)\to -{1\over 16}\big(kr_0\big)^{-({1\over
t}-1)} t^{{\sigma\over 2}-4}{1\over kR}\big({r_2\over R}\big)^{{\sigma\over
2}-3}  {\rm e}^{-{R\over 2r_0}({1\over t}-1)}.  \eqno(25) $$                  
    The integrability of this expression in $t=(0,1)$  is by Eq.(22) obvious.                                                           
                  
\medskip 
\ \ \  \ \ {\bf   Case PEE} 
 \smallskip     
This class  should lie near the pure supersingular case, 
where $g^2={\rm fixed}$. The potential  is introduced as                     
      $$ g^2U(s;r)= {1\over r_0^2} \big({r_0\over
R}\big)^{\sigma_0}{\rm 
e}^{r_1s\over r} {\rm e}^{-{r\over r_2}}. \eqno(26) $$
The large-$R$ form of the Master equation (2) reads so                        
     $$ {\rm e}^{r_1s(R)\over R}\to k^2r_0^2\big({R\over
r_0}\big)^{\sigma_0}{\rm e}^{R\over r_2}, \ \  [R\to\infty], \eqno(27) $$     
                         which suggests the relationship 
O$\{s(R)\}>$O$\{{R^2\over r_1r_2}\}$.  Incorporation of the definition (27)
into Eq.(11) yields in the region $\epsilon$                               
                                    $$ K^2_\epsilon(t)\to
k^2\big(k^2r_0^2\big)^{{1\over t}-1}\big({R\over r_0}\big)^{\sigma_0({1\over
t}-1)}{\rm e}^{{R\over r_2}({1\over t}-t)}, \ \ [R\to\infty]. \eqno(28) $$   
This expression furnishes at $R\to \infty$ the discriminant (16)  as         
              $$ p_\epsilon(t)\to -{1\over 16}\big({1\over t^2}+1\big)^2
{1\over  kR}\big( {R\over r_0}\big)^2\big(kr_0\big)^{1-{1\over
t}}\big({r_0\over R}\big)^{{1\over 2}\sigma_0({1\over t}-1)}{\rm e}^{-{R\over
2r_2}({1\over t}-t)}. \eqno(29) $$                                            
            The exponential factor involved guarantees  the existence of 
$P_\epsilon(t)$ in $t=(0,1)$.                                                                         

\medskip
\ \ \ \ \  {\bf   Case PEP }
\smallskip                       
               We shall now  treat the scattering potentials                  
            $$ g^2U(s;r)={1\over r_0^2}\big({r_0\over R}\big)^{\sigma_0}{\rm
e}^{r_1s\over r} \big({r_2\over r_2+r}\big)^{\sigma_2},\ \ \
[\sigma_2>8].\eqno(30) $$                                                    
The corresponding Master equation (2) becomes asymptotically                  
      $$  {\rm e}^{r_1s(R)\over R}\to k^2r_0^2\big({R\over
r_0}\big)^{\sigma_0}\big({R\over r_2}\big)^{\sigma_2},\ \ [R\to\infty].
\eqno(31) $$                                                                
The order relationship  between the parameters is extracted 
now as 
O$\{s(R)\}>$ O$\{{R\over r_1}\}$. The local wavenumber square (11)  
and the discriminant (16) are  therefore obtained in the 
region $\epsilon$ as                
            $$ K_\epsilon^2(t)\to k^2
\big[ k^2r_0^2 {R^{\sigma_0+\sigma_2}\over r_0^{\sigma_0}r_2^   
{\sigma_2}} \big] ^{{1\over t}-1}{1\over t^{\sigma_2}},\ \
[R\to\infty],\eqno(32)$$                                                  
           and                     $$p_\epsilon\to -{1\over
16}t^{{\sigma_2\over 2}-4}{1\over kR}\ln\big({R^{\sigma_0+\sigma_2}\over
r_0^{\sigma_0}r_2^{\sigma_2}}\big)\big[{R^{\sigma_0+\sigma_2}\over
r_0^{\sigma_0}r_2^{\sigma_2}}\big]^{-{1\over 2}({1\over t}-1)}. \eqno(33) $$
The integrability of the last expression is by analysis obvious both at fixed 
and increasing values of $R$  as well as near and off the origin $t=0$.      
                                                                    
\medskip 
\ \ \  \ \  {\bf    Case EPE }  
\smallskip                                                                                                          
    Off the singularity region this potential class develops in our double
limit the strongest suppression. Indeed,                                      
          $$ g^2U(s;r)={1\over r_0^2} {\rm e}^{-{R\over r_0}}\big( {r_1+r\over
r}\big)^s{\rm e}^{-{r\over r_2}}. \eqno(34) $$                              
The Master equation (2) can now be recast for asymptotical parameter values as
     $$ {\rm e}^{r_1s(R)\over R}\to k^2r_0^2 {\rm e }^{R({1\over r_0}+{1\over
r_2})}, \ \  [R\to\infty]. \eqno(35) $$                                     
The singularity parameter $s(R)$ increases thus  proportionally to  $R^2$.    
Incorporation of  the expression (35) into Eq. (11) yields for the exponential
region            $$ K_\epsilon^2(t)\to{1\over r_0^2}\big[k^2r_0^2{\rm
e}^{R({1\over r_0}+{1\over r_2})}\big]^{1\over t}{\rm e}^{-R({1\over
r_0}+{t\over r_2})}, \ \ [R\to\infty]. \eqno (36) $$    The discriminant  
(16) is extracted hence as       $$ p_\epsilon(t)\to -{1\over 16}\big({1\over
r_0}+{1\over r_2}\big)^2\big( {r_0R\over t^4}\big) {\rm e}^{-{1\over
2}R({1\over r_0}+{1\over r_2}){1\over t}}, \ \  [R\to\infty]. \eqno (37) $$   
            Near  the singularity point $t=0$, the exponential decrease of the
potential  at  $t={\rm fixed}$  dominates for \ \ $R\to \infty$ the powerlaw
increase there. The quantity $P_\epsilon(t)$  is thus finite.    
\medskip  
\ \ \ \ \ {\bf   Case EPP}  
 \smallskip       
This interaction is very similar to

the previous one within the region near  the singularity point. The  potential
reads  now   
      $$ g^2U(s;r)= {1\over r_0^2} {\rm e}^{-{R\over
r_0}}\big({r_1+r\over r}\big)^s\big({r_2\over r_2+r}\big)^\sigma, \ \ 
[\sigma>2]. \eqno(38) $$  The Master equation (2) governs the large-$R$
dependence of the singularity parameter  $s$  as 
         $$ {\rm e}^{r_1s(R)\over R}\to
k^2r_0^2\big({R\over r_2}\big)^\sigma,\ \ [R\to\infty]. \eqno(39) $$   
Hence one extracts the order relationship 
${\rm O}\{s(R)\}> {\rm O}\{{R\over r_1}\}$ for $R\to \infty$. On the other
hand, the asymptotical forms of (11) and (16) follow from  relationship (39) as
                    $$ K^2_\epsilon(t)\to k^2{1\over
{t^\sigma}}\big[k^2r_0^2{\rm e}^ {{ R\over r_0}} \big({R\over
r_2})^\sigma\big]^{{1\over t}-1},\ \ [R\to\infty],\eqno(40) $$                
as well as the simultaneously and exponentially decreasing discriminant       
                                          $$ p_\epsilon(t)\to -{1\over
16}{R\over r_0^2k^2 t^4}{\rm e}^{-{R\over r_0}({1\over t}-1)}, \ \ 
[R\to\infty]. \eqno(41) $$     This again means convergence of the
corresponding series  (5).                                             
\bigskip
\bigskip  
\ \ \ \ \  {\bf   Case PPP }   
\smallskip                     
The interaction behind this symbol is written in the variable $r$ as         
      $$ g^2U(s;r)={1\over r_0^2}\big({r_0\over
R}\big)^{\sigma_0}\big({r_1+r\over r}\big)^s\big({r_2\over
r_2+r}\big)^{\sigma_2},\ \  [\sigma_2>4]. \eqno(42) $$                        
The asymptotical Master equation is  now extracted  as                        
 $$ {\rm e}^{r_1s(R)\over R}\to k^2r_0^2\big({R\over
r_0}\big)^{\sigma_0}\big({R\over r_2}\big)^{\sigma_2},\ \
[R\to\infty].\eqno(43) $$                                                   
The increase of $s(R)$  for large values of the matching distance is  slightly
more rapid than that of ${R\over r_1}$.  The local wave number square  becomes
in our double limit                                                           
  $$ K^2_\epsilon(t)\to k^2{1\over t^{\sigma_2}}\big[k^2r_0^2\big({R\over
r_0}\big)^{\sigma_0}\big({R\over r_2}\big)^{\sigma_2}\big]^{{1\over t}-1},    
\ \,\eqno(44) $$      while the discriminant (16) develops then the
asymptotical form                                   $$ p_\epsilon (t)\to
-{1\over 16}\ln\big({R^{\sigma_0+\sigma_2}\over
r_0^{\sigma_0}r_2^{\sigma_2}}\big){1\over kR}t^{{1\over 2}(\sigma_2-4)}
\big[k^2r_0^2{R^{\sigma_0+\sigma_2}\over
r_0^{\sigma_0}r_2^{\sigma_2}}\big]^{-{1\over 2}({1\over t}-1)}.\eqno(45) $$   
                                          The exponential decay involved in
the limit  $R\to\infty$  for  $t<1$  ensures the existence of $P_\epsilon(t)$
in region $\epsilon$. 
\medskip 
\ \ \  \ \    {\bf   Case PPE}   
\smallskip    
                                                                            
The core factor implies, in fact, for the double limit we are interested in, 
again a hidden  exponential dependence on the singularity parameter $s$ .
Indeed,                                                                       
        $$ g^2U(s;r)= {1\over r_0^2}\big({r_0\over R}\big)^{\sigma_0}
\big({r_1+r\over r}\big)^s{\rm e}^{-{r\over r_2}}.\eqno(46) $$               
By analysis, one obtains the Master equation in the form                      
       $$ {\rm e}^{r_1s(R) \over R} \to k^2r_0^2\big( {R \over
r_0}\big)^{\sigma_0}{\rm e}^{R\over r_2},\ \ [R\to\infty].\eqno(47) $$  
This result implies O$\{s(R)\}>$O$\{{R^2\over r_1r_2}\}$. One  also  concludes
from the asymptotical expression  (47) that                                    
               $$ K_\epsilon^2(t)\to
k^2\big[k^2r_0^2\big({R\over r_0}\big)^{\sigma_0}\big]^{{1\over t}-1}{\rm
e}^{{R\over r_2}({1\over t}-t)}, \ \ [R\to\infty], \eqno(48) $$               
                                and                               $$
p_\epsilon(t)\to -{1\over 16} {1\over kR}\big({R\over r_2}\big)^2\big({1\over
t^2}+1\big)^2{\rm e}^{-{R\over r_2}({1\over t}-t)}, \ \ [R\to\infty].
\eqno(49) $$   Except for the  single point $t=1$, this expression 
exponentially vanishes    in the limit $R\to\infty$  within  the region 
$\epsilon$. Consequently, the function $P_{\epsilon}(t)$ of  the identity (15)
does exist  also in this case.   \medskip
 The behaviour of the functions $p_\gamma(t)$, $[\gamma=\epsilon,\tau]$,  
of (16) have  been studied for eight classes of repulsive  singular
potentials along both of the complementary regions $\epsilon$  and  $\tau$.
Owing to the existence and boundedness of the functions  $P_\gamma(t)$       
of (15),  both series (5) and (6) are  absolutely convergent and reduce in the
limit we are interested in  to the respective leading terms [1,2].
Accordingly, the scattering wave functions develop for  $R\to\infty$ the
following asymptotical forms                      $$ u^+(t)\to \big({k^2\over
K_ \epsilon^2(t)}\big) ^{1\over 4}  {\rm e}^{R\int_1^t{\rm
d}t'|K_\epsilon(t')|},\ \ [t<1],\eqno(50) $$ as well as
       $$ u^+(t)\to \big( {k^2\over
K^2_\tau(t)} \big) ^{1\over 4} \big\{ C^+_0\cos [R\int_1^t {\rm
d}t'K_\tau(t')] + S^+_0 \sin [R\int_1^t {\rm d}t'K_\tau(t')] \big\}, \ \
[t>1].\eqno(51) $$    The constants $C^+_0$  and $S^+_0$ involved  are,
in fact, $R$-dependent and fixed uniquely  by postulating smooth matching
of the external wave function (51) to (50), the internal one, at
$t=1$. The functions  $K_\gamma(t)$ are supplied  for inclusion into (50)-(51)
 by  the large-$R$ expressions  (19), (24), (28), (32), (36), (40), (44) and
(48)  for the respective potential classes.  \medskip    The overall
conclusions extracted  can be lumped into the following three points: (a) The
double limits of combining vanishing linear and diverging nonlinear potential
parameters may give rise, at different types of  interdependence of these
variable constants, to reasonable scattering problems, (b) These  are solved 
by  pairs of absolutely convergent series, (c) the lengths of which get
reduced, just in the limit scrutinized, to single terms  calculable by
quadrature.    \smallskip     Finally, only  a slight hint at philosophy. The
coupling constant         $g^2$ may be regarded as the $quantity$ inherent in
the singularity of the scattering potential, while the stage $s$  of
singularity could be a measure of its $quality$. The above argument may thus 
yield an example for treating the relationship  between  'quality' and
'quantity'  under extreme circumstances.             
 \bigskip  
{\bf Acknowledgement}
Many thanks are due to my colleagues, including Dr. G. Bencze and Dr. I. Racz,
for valuable discussions. The work was supported by the Hungarian NSF under
Grant No. OTKA 00157.   \bigskip  
{\bf References}  
\bigskip
[1] \ \  T.Dolinszky: J.Math.Phys. {\bf 36}, 1621(1995) 
\hfill \break  
[2] \ \  T.Dolinszky: J.Math.Phys. {\bf 38}, 16 (1997) 
\hfill \break 
[3] \ \  T.Dolinszky: Supersingular scattering:  math-ph./0002047 24 Feb 2000

\vfill
\end